\documentclass[superscriptaddress,amsmath,amssymb,onecolumn,nofootinbib]{revtex4}

\usepackage{dcolumn}
\usepackage{bm}
\usepackage{ulem}
\usepackage{xfrac}
\usepackage{lipsum}
\usepackage{braket}
\usepackage{dsfont}
\usepackage{phoenician}
\usepackage{mathtools}
\usepackage{slashed}

\usepackage[printwatermark]{xwatermark}
\usepackage{xcolor}

\usepackage{graphicx,amsmath,amssymb}
\usepackage{epstopdf}
\usepackage[colorlinks=true, urlcolor=blue, linkcolor=blue, citecolor=blue]{hyperref}

\begin{document}

\begin{flushright}
{\small Preprint No.\ JLAB-PHY-26-4808
}
\end{flushright}

\title{Insight on confinement from the QCD effective charge}

\author{A.~Deur} 
\affiliation{Thomas Jefferson National Accelerator Facility, Newport News, Virginia 23606, USA}


\date{\today}


\begin{abstract}
Like the Gell-Mann--Low coupling $\alpha$ of QED, the effective charge $\alpha_{g_1}$ is a physical, observable-defined running coupling that can serve as the QCD coupling $\alpha_s$. It can be represented by an analytic form consistent with renormalization-group evolution, holographic light-front QCD (HLFQCD), and the world data on $\alpha_s$. As an observable,  $\alpha_{g_1}$ has a physically relevant analytic structure that reflects the underlying partonic dynamics. It displays, in the long-distance regime, two imaginary conjugate singularities at $Q^2 = \pm i\Lambda_s^2$, with $\Lambda_s$ the QCD scale. To connect this structure to confinement, 
 we combine known general relations between $\alpha_{g_1}$ and the parton dressed propagators and vertices with a recent phenomenological representation of $\alpha_{g_1}$ within HLFQCD.
Since, unlike vertices, propagators characterize a field variation between separate locations, the two singularities are assigned to the propagators. This results in the parton propagators displaying a long-distance behavior $e^{-\Lambda_s |x|/\sqrt{2}}  |x|^{-5/2 + d_a}$ where propagation is suppressed beyond distances of order $1/\Lambda_s$, as expected from confinement. This offers an intuitive interpretation of confinement as the suppression of QCD Green's functions at long distance.

\end{abstract}

\maketitle

\section{Introduction\label{intro}}

Understanding quark confinement from the basic equations of quantum chromodynamics (QCD) remains a problem under active investigation~\cite{Gross:2022hyw}.
Recent studies of the QCD coupling over all momentum 
scales~\cite{Brodsky:2010ur, deTeramond:2024ikl, deTeramond:2025qlj} have uncovered an interesting phenomenon: 
as the momentum scale $Q^2$ decreases relative to a characteristic non-perturbative scale, e.g. the nucleon mass, the Landau 
pole appearing in the perturbative coupling flows in the complex $Q^2$ plane from its initial position on the real axis to the imaginary axis. Here, we study how this phenomenon connects to confinement.

In perturbative QCD (pQCD), the coupling $\alpha_s(Q^2)$ has a Landau pole located at $Q^2 \simeq \Lambda_s^2$, where $Q$ is
the characteristic momentum of a reaction. 
The pole is unphysical for several reasons~\cite{Deur:2016tte,Deur:2023dzc}, including its violation of the expected analyticity properties of observables. 
In fact, no measurement has revealed any hint  of special behavior, divergence, cusp, etc. near $Q^2\simeq \Lambda_s^2$. In particular,
a QCD effective charge~\cite{Grunberg:1980ja,Grunberg:1982fw} has no Landau pole~\cite{Deur:2005cf}.
This makes effective charges a convenient definition of $\alpha_s$, which we adopt here. 
Then, $\alpha_s$ is an observable quantity and findings based on it are objective, 
in contrast to other definitions of $\alpha_s$ for which statements can be renormalization scheme (RS) and gauge dependent. 
Furthermore, it enables the powerful framework of analyticity~\cite{Pascalutsa:2024hvy}.  A particularly suitable effective charge is that obtained from
the Bjorken sum rule~\cite{Bjorken:1966jh, Bjorken:1969mm}, a fundamental quantum field theory (QFT) relation derived for $Q^2\to \infty$. 
Its extension to finite $Q^2$ has been calculated with pQCD~\cite{Kataev:1994gd,Kataev:2005hv,Baikov:2010je} 
and agrees well with the data~\cite{SpinMuon:1993gcv, SpinMuonSMC:1994met, SpinMuonSMC:1997voo, SpinMuon:1995svc, SpinMuonSMC:1997mkb, COMPASS:2010wkz, HERMES:1998pau, HERMES:2000apm, HERMES:2002gmr, Deur:2004ti, Deur:2008ej, Deur:2014vea, Deur:2021klh, E143:1994vcg, E143:1995rkd, E142:1996thl, E143:1995clm, E143:1996vck, E154:1997xfa, E154:1997ysl, E143:1998hbs, E155:1999pwm, E155:2000qdr}.
Denoting $g_1^{p(n)}$ the proton (neutron) first spin structure function~\cite{Deur:2018roz}, $x_{\rm Bj}$ the Bjorken scaling variable~\cite{Bjorken:1968dy}, 
$g_A$ the nucleon axial charge and $\alpha_s^{\rm pQCD}$, the pQCD coupling used for the pQCD corrections~\cite{Kataev:1994gd,Kataev:2005hv,Baikov:2010je}, 
the Bjorken sum rule pQCD approximant is
\begin{equation}
\Gamma_1^{p-n}(Q^2) \equiv \int_0^1 \left[g_1^p(x_{\rm Bj},Q^2) - g_1^n(x_{\rm Bj},Q^2)\right] dx_{\rm Bj} = \frac{g_A}{6}
\left[ 1 - \frac{\alpha_s^{\rm pQCD}(Q^2)}{\pi} + \mathcal O \left( \big(\alpha_s^{\rm pQCD}\big)^2 \right) + \mathcal O\left(\frac{1}{Q^2}\right) \right].  
\label{eq:BJSR}
\end{equation}
From it, the effective charge $\alpha_{g_1}(Q^2)$ is defined~\cite{Grunberg:1980ja,Grunberg:1982fw} as $\Gamma_1^{p-n}(Q^2) \equiv \frac{g_A}{6} \left[ 1 - \frac{\alpha_{g_1}(Q^2)}{\pi} \right]$,
viz.
\begin{equation}
\alpha_{g_1}(Q^2) = \pi \left[1 - \frac{6}{g_A}\Gamma_1^{p-n}(Q^2)\right], 
\label{eq:ag1_def}
\end{equation}
where the $\mathcal O \big( (\alpha_s^{\rm pQCD})^2 \big)$ and $\mathcal O(1/Q^2)$ corrections became folded in the $Q^2$-dependence of 
$\alpha_{g_1}$. Eq.~(\ref{eq:ag1_def}) holds for any $Q^2 \geq 0$ value. It is a convenient effective charge because one may identify 
$\alpha_{g_1}$ to the QCD coupling at all $Q^2$~\cite{Deur:2005rp, Deur:2016tte}. 
Moreover, as an observable, $\alpha_{g_1}$ obeys the general QFT principles of causality and unitarity, 
which imply its analyticity in the complex $Q^2$ plane. Finally, $\alpha_{g_1}$ is well measured from high $Q^2$ (ultraviolet, UV) to low $Q^2$ (infrared, IR)~\cite{Deur:2005cf,Deur:2008rf,Deur:2022msf} and 
is consistent with derivations from AdS/QCD~\cite{Brodsky:2010ur, deTeramond:2024ikl, deTeramond:2025qlj} using its Holographic Light-Front QCD (HLFQCD) 
implementation~\cite{Brodsky:2006uqa,Brodsky:2007hb,deTeramond:2008ht,Brodsky:2014yha}, and from Lattice QCD/Dyson-Schwinger 
equations (DSE)~\cite{Binosi:2016nme,Cui:2019dwv}.

\begin{figure}[h] 
\includegraphics[width=6cm]{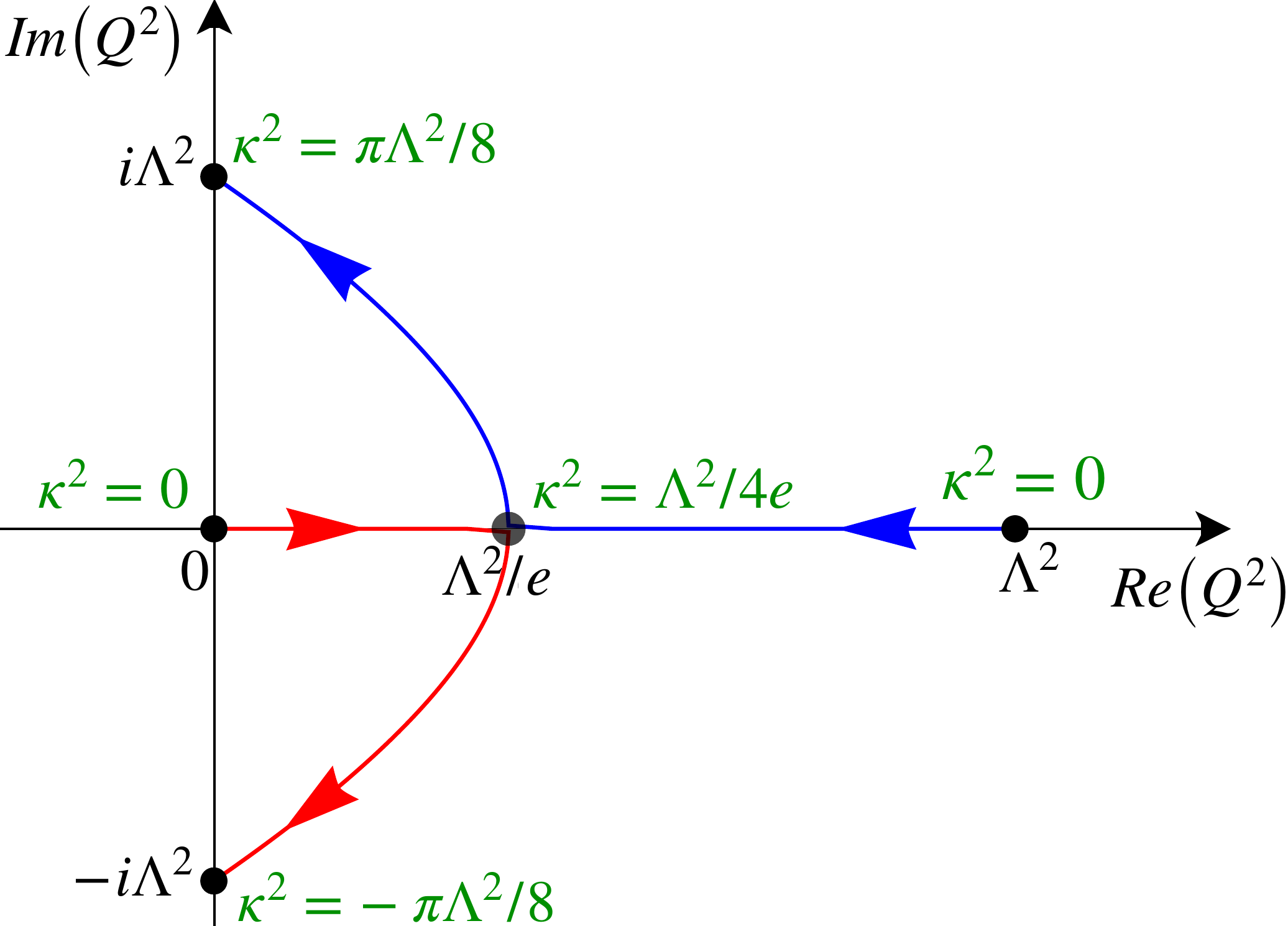}
\caption{Flow of the $\alpha_{g_1}$ poles in the complex $Q^2$-plane with the variation of the HLFQCD scale $\kappa^2$. (Adapted from~\cite{deTeramond:2024ikl}).
\label{fig:flow}}
\end{figure}

The Landau pole of $\alpha_s^{\rm pQCD}$ signals both a breakdown of perturbation theory and a violation of
analyticity. Since in the UV, $\alpha_{g_1} = \alpha_s^{\rm pQCD} + \mathcal O \big( (\alpha_s^{\rm pQCD})^2 \big)$, the pQCD expression of $\alpha_{g_1}$ also displays the pole. 
Yet, the full $\alpha_{g_1}$, as measured, has no pole and indeed,  
extending $\alpha_{g_1}$ into the IR domain using HLFQCD eliminates the pole~\cite{Brodsky:2010ur, deTeramond:2024ikl}. 
One may track how this occurs by 
evolving the HLFQCD nonperturbative scale $\kappa$ from 0 to its physical value around half the nucleon mass, $\kappa \simeq M_N/2$~\cite{Brodsky:2014yha}, see Fig.~\ref{fig:flow}. 
Setting $\kappa=0$ corresponds to neglecting it in front of $Q^2$, i.e. to be in the pQCD domain. Two poles lie on the real axis, at $Q^2=\Lambda_s^2$ (the Landau pole) and at $Q^2=0$. Formally, the latter comes from $1/[u \ln(\sfrac{u}{\Lambda_s^2})]$ in Eq.~(1) of~\cite{deTeramond:2024ikl} (here, the RHS of Eq.~(\ref{eq:alpha_HLF})) and is imposed by dimensional consistency of the denominator $4\kappa^2 +u \ln(\sfrac{u}{\Lambda_s^2})$.  
As $\kappa$ increases, the poles move on the real axis and when $\kappa^2 \to \Lambda_s^2/(4e)$, they merge at $Q^2=\Lambda_s^2/e$. 
For $\kappa^2 >\Lambda_s^2/(4e)$ they split into a complex conjugate pair and flow in the complex plane until they reach 
pure imaginary values $Q^2=\pm i\Lambda_s^2$ when $\kappa^2 = \frac{\pi}{8}\Lambda_s^2$.

$\kappa^2 = \frac{\pi}{8}\Lambda_s^2$  corresponds to a configuration of maximal analyticity, viz the poles are maximally imaginary. 
It directly influences physical amplitudes since real poles characterize particle-like excitations, 
whereas imaginary poles correspond to dissipative modes with no resonant structure, 
and thus, do not represent asymptotic or even metastable states. 
In fact, complex conjugate poles are generally associated with suppression of a system evolution, be it classical or quantum~\cite{Arfken-Weber}.
This suggests that the value of $\Lambda_s$, or equivalently $\kappa$, is set so that the poles of $\alpha_s$ 
are maximally imaginary, thereby suppressing the spatial propagation of color particles.
This is supported by the remarkable fact that $\kappa^2 = \frac{\pi}{8}\Lambda_s^2$ is fulfilled~\cite{deTeramond:2024ikl} by the values of $\kappa$ and $\Lambda_s$
despite them stemming from different phenomenologies (hadron mass spectra for $\kappa$~\cite{Brodsky:2014yha}, and pQCD approximants matched to 
corresponding UV data for $\Lambda_s$~\cite{Deur:2023dzc}).
In this perspective, confinement becomes strictly connected to the analytic structure of $\alpha_s$: the lack of real poles,
their flow to purely imaginary ones, and their maximal separation prevent asymptotic colored states.

The pole flow shown in Fig.~\ref{fig:flow} thus suggests an interesting heuristic interpretation of the transition between perturbative and confined QCD dynamics.
While $\kappa$ is fixed by nature, it can still be varied to see to what physical situation such value would correspond. 
For example, setting $\kappa=0$ is equivalent to it being fully negligible compared to the relevant scale of the reaction, viz $Q^2$. 
So the fictitious $\kappa=0$ case corresponds to the physical pQCD $Q^2 \to \infty$ case. 
Fig.~\ref{fig:flow} shows that the poles are real until $\kappa^2=\kappa_{\rm crit}^2=\Lambda_s^2/4e$. Then maximum suppression of propagation occurs at 
$\kappa_{\rm conf}^2=\pi \Lambda_s^2/8$. As long as the poles are real, propagation occurs freely and the system follows a pQCD behavior. Then, the confinement onset starts at $\kappa_{\rm crit}$. The scale characterizing a dynamical domain may then be quantified by $Q_*^2=\Lambda_s^2 \kappa_{\rm conf}^2/\kappa_*^2$, with $\kappa_*$ the value of the $\kappa$ parameter. For confinement, $Q_*^2 = \Lambda_s^2$. 
The case $Q_*^2 = \Lambda_s^2\kappa_{\rm conf}^2/\kappa_{\rm crit}^2 = 3.1$ GeV$^2$ ($\Lambda_s=0.85$ GeV in the $g_1$ RS) indicates the start of the confinement onset. 
In all, between the Bjorken limit, $\kappa_*=0$ (viz $Q_*^2 \to \infty$) until $Q_{\rm crit}^2 =3.1$ GeV$^2$ the system behaves as asymptotically free. Then confinement is essentially realized at $Q_*^2 = \Lambda_s^2 \kappa_{\rm conf}^2/\kappa_{\rm conf}^2=0.72$ GeV$^2$. 
Thus, the pole flow quantifies the momentum scales characterizing the different dynamical domains of QCD, from asymptotic freedom to confined behavior.

The analytic structure of $\alpha_{g_1}$ reflects those of the underlying correlation functions~\cite{Huber:2018ned}.
The emergence of complex poles is therefore consistent with the lack of real poles in quark and gluon propagators, as expected from confinement although
here, confinement is not imposed but arises from the analytic structure of the theory once a scale is generated. 
In this article, we study how the $\alpha_{g_1}$ poles constrain
the correlation function singularities. 
It provides an intuitive dynamical understanding of confinement. Just like imaginary poles in classical settings dissipate energy and damp
the evolution of a mechanical or electrical system,  the imaginary poles of $\alpha_{g_1}$ damp the propagation of partonic fields once those are
far from their emission point.   

\section{Analytic structure of parton propagators}

In the effective charge framework, choosing an observable is equivalent to a RS choice~\cite{Brodsky:1994eh, Deur:2016cxb}.
 In particular,  $\alpha_{g_1}$  defines the $g_1$ RS. 
In any RS, $\alpha_s$ can be expressed in terms of dressing functions (DF) $Z_i$ 
renormalizing the bare QCD Green's functions~\cite{Deur:2016tte,Deur:2023dzc,Fischer:2006ub}. Taking e.g., the quark-gluon interaction, 
\begin{equation}
\alpha_{g_1}(Q^2) = \alpha_{g_1}(Q_0^2) Z_1^2(Q^2)Z_2^2(Q^2) Z_3(Q^2),
\label{eq:a_s-qqg-vert}
\end{equation}
where $Z_1$, $Z_2$, and $Z_3$ denote the DFs of the quark-gluon vertex, the quark propagator and of the gluon propagator, respectively. 
Besides $Q^2$, the $Z_i$ also depend on a reference scale $Q_0^2$ but unless useful, we keep it implicit to lighten notations. Likewise, 
we do not label that they are in the $g_1$ scheme nor that they can be gauge dependent (as are the bare parton propagators).
Similar expressions stem from the 3-gluon interaction,
\begin{equation}
\label{eq:a_s-3g-vert}
\alpha_{g_1}(Q^2) = \alpha_{g_1}(Q_0^2)Z_3^3(Q^2)Z^2_{3g}(Q^2),
\end{equation}
where $Z_{3g}$ is the 3-gluon vertex DF, and from the 4-gluon interaction. The latter will not be discussed here because
$\alpha_{g_1}$ defines in coordinate space an effective interaction 
$F(r) =\alpha_{g_1}(r)/r^2$,
where the $1/r^2$ comes from one-gluon exchange. All other $r$-dependent effects, mainly from gluon self-interaction, are folded in 
$\alpha_{g_1}(r)$~\cite{Deur:2023dzc, Brodsky:2024zev}.
This definition selects interactions consistent with a force between two color charges. 
Quark-gluon and 3-gluon interactions are compatible with such system but not the 4-gluon interaction.
Since it is unclear whether a coupling obtained from the latter is compatible with $F(r)$, we will not consider it.
Likewise, we will not  consider the relation involving the ghost-gluon vertex (Taylor coupling~\cite{Deur:2023dzc}) since it is associated with the ghost fields used in covariant-gauge QCD, whereas the light-front formulation employed in HLFQCD does not introduce ghost fields.

Given Eqs.~(\ref{eq:a_s-qqg-vert}) and~(\ref{eq:a_s-3g-vert}), $\alpha_{g_1}$ and the $Z_i$ combinations must have corresponding singularities. 
One naturally assigns the singularities of $\alpha_{g_1}$ to propagators rather than to vertices since confinement is
a feature of propagation, viz $Z_2$ and/or $Z_3$ convey the singular structure while $Z_1$ and 
$Z_{3g}$ are analytic. 
In fact, we show in Section~\ref{Constraints on  b_i} of the Supplemental Material that if $Z_1$ is analytic, then so is $Z_{3g}$. 
Thus, here, we need assume only that one vertex function is analytical. 
Furthermore, lattice QCD has long established that 
confinement already occurs in the pure glue sector~\cite{Gross:2022hyw}, and that static quarks are confined. This shows that the main IR 
mechanism driving confinement is gluonic rather than due to singular quark dynamics/quark-gluon vertex, which also motivates $Z_1$ being analytic.
Within this picture, it is then natural for the dominant nonanalytic structure to be in the propagators. 
Actually, would $Z_1$ be non-analytical, then $Z_{3g}$ would need to be fine-tuned so that their singularities cancel (Section~\ref{Constraints on  b_i}). 
Otherwise, the remaining non-analytical structure would conflict with $\alpha_{g_1}$ being an observable. 
Finally, that the deviation from the classical $F(r)  \propto 1/r^2$ law is dominated by IR 
processes, while vertices are point-like quantities, also supports this choice.

\subsection{Propagation and Confinement \label{Prop-Conf}}

The quark and gluon dressed propagators are
\begin{equation}
F(Q^2) = Z_2(Q^2)  \frac{\slashed{Q} +M(Q^2)}{Q^2 - M^2(Q^2) + i\epsilon}, 
\qquad
D^{ab}_{\mu\nu}(Q^2) =   \delta^{ab} \frac{Z_3(Q^2)}{Q^2 + i\epsilon} \left[ g_{\mu\nu} + \Gamma_{\mu \nu} \right],
\nonumber
\end{equation}
respectively, 
where $\Gamma_{\mu \nu}$ in the gluon propagator is a gauge-dependent term (e.g., $\Gamma_{\mu \nu} = (1 - \xi) \frac{q_\mu q_\nu}{Q^2}$ 
in covariant linear gauges, with $\xi$ the gauge parameter, 
or $\Gamma_{\mu \nu} =- \frac{n_\mu q_\nu + n_\nu q_\mu}{n \cdot q}$ in the Light Cone gauge),
$a,b$ color indices (to keep notations light, they will not be included anymore), 
and $M(Q^2)$ is the quark mass function. 
We will neglect, in the context of HLFQCD, the $Q^2$ dependence of $M(Q^2)$ and set $M(Q^2)=m$, the light current-quark mass. 
Indeed, in the HLFQCD framework, the only emergent scale parameter that breaks conformal symmetry is $\kappa$, 
which characterizes the confinement potential and thus, the gluons.  
Furthermore, this work pertains to the light quark sector for which Eq.~(\ref{eq:BJSR}) stands, 
and $\alpha_{\rm eff}^{\rm HLF}, $ Eq.~(\ref{eq:alpha_HLF}), is applicable. Here $m$ is negligible, $m \ll \Lambda_s$. 
Thus, the mechanism leading to a large constituent quark mass in other frameworks, e.g., lattice QCD 
or DSE approaches~\cite{Ding:2022ows} manifests differently in the HLFQCD approach where quark masses are either neglected (light quark sector) or 
taken as the current quark mass~\cite{Brodsky:2014yha, deTeramond:2025qlj}. Likewise, no gluon effective mass appears explicitly in HLFQCD. 
Thus, setting $M(Q^2)=m$ underlines the absence of an explicit dynamical mass generation mechanism in the HLFQCD framework used here rather than implying 
a general momentum independence of quark mass functions.
This simplifies the quark dressed propagator to
\begin{equation}
F(Q^2) = Z_2(Q^2)  \frac{\slashed{Q} + m}{Q^2 - m^2 + i\epsilon}.
\label{eq:dressed-prop-quarks}
\end{equation}
Furthermore, since $\alpha_{g_1}$ is an observable, its analytic structure is gauge independent. 
Thus, we suppress the gauge-dependent term $\Gamma_{\mu \nu}$ in the gluon propagator and keep only the scalar propagator factor,
\begin{equation}
D(Q^2) = \frac{Z_3(Q^2)}{Q^2+ i\epsilon},
\label{eq:dressed-prop-gluons}
\end{equation}
which should be understood as a propagator-like factor entering the analytic representation of $\alpha_{g_1}(Q^2)$, Eq. (\ref{eq:alpha_HLF}), 
rather than as a calculation of a particular gauge-fixed gluon Green function.

To determine the analytic structure of the parton DFs, we equate Eq.~(\ref{eq:a_s-qqg-vert}) to the HLFQCD expression 
for $\alpha_{g_1}$ from~\cite{deTeramond:2024ikl}, which captures its correct UV and IR behaviors: 
\begin{equation}
\alpha_{g_1}(Q^2) = \alpha_{g_1}(Q_0^2)Z_1^2(Q^2,Q_0^2) Z_2^2(Q^2,Q_0^2)Z_3(Q^2,Q_0^2)   = \alpha^{\rm HLF}_{\rm eff}(Q^2) = \pi \exp\left[-\int_0^{Q^2} \frac{d u}{4 \kappa^2 + u \ln\big(\frac{ u}{\Lambda_s^2}\big)} \right],
\label{eq:alpha_HLF}
\end{equation}
where the RHS is independent of $Q^2_0$.
 It is worth noting that the RHS of Eq.~(\ref{eq:alpha_HLF}) is a phenomenological representation of $\alpha_{g_1}(Q^2)$, 
rather than an expression obtained directly from QCD Lagrangian. It accurately describes the measured $\alpha_{g_1}(Q^2)$ while incorporating 
both the HLFQCD IR prediction and the pQCD UV behavior. 
In the analysis below, the phenomenological identification of the analytic factors on the RHS of Eq.~(\ref{eq:alpha_HLF}) with the 
LHS DFs is a phenomenological construction that need not yield the familiar QCD propagators extensively studied with DSE, lattice QCD, and related continuum methods~\cite{Huber:2018ned,Fischer:2006ub}. Rather, it investigates whether the analytic structure empirically encoded in $\alpha_{g_1}(Q^2)$ 
can be represented by effective propagator-like functions, and the physical consequences of such a representation. 

\begin{figure}[h] 
\includegraphics[width=7cm]{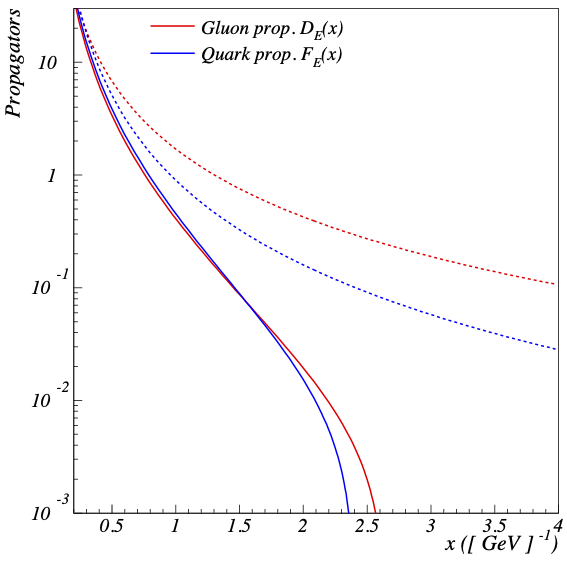}
\caption{Gluon (red line, $D_E(x)$) and quark (blue line, $F_E(x)$) Euclidean propagators vs. distance $x$ in the asymptotic large $x$ approximation.
The dashed lines are for the free massless boson case ($D_E \propto x^{-2}$), 
and massive fermion case ($F_E(x) \propto e^{-mx}(x^{-5/2} + mx^{-3/2})$, with $m=5$~MeV, 
to contrast with the confined propagation $\propto e^{-\frac{\Lambda_s}{\sqrt{2}}x}/x^{5/2-d_a}$. 
The propagators are normalized to the same value at $x=0.2$ GeV$^{-1}$ 
and we set to zero the constant parts of the phases $\theta(|x|)$ and $\theta'(|x|)$. 
Confinement is reflected by the exponential damping stemming from the imaginary singularities, while $d_a \simeq 0.1$ gives a small anomalous power correction coming from the same singularities.
}
\label{fig:prop}
\end{figure}

The integrand in Eq.~(\ref{eq:alpha_HLF}) has poles for $4\kappa^2 + u \ln\big(\frac{u}{\Lambda_s^2}\big) = 0$, discussed in Section~\ref{intro}.    
Eq.~(\ref{eq:alpha_HLF}) can be rewritten as
\begin{equation}
2 \ln Z_1(Q^2) + 2 \ln Z_2(Q^2) + \ln Z_3(Q^2)  + C = I(Q^2),
\label{eq:constraint}
\end{equation}
where
\begin{equation}
I(Q^2) \equiv - \int_0^{Q^2} \frac{du}{4\kappa^2 + u \ln\big(\frac{u}{\Lambda_s^2}\big)} = \ln\Big(\frac{\alpha_{g_1}(Q^2)}{\pi}\Big), 
\label{eq:I(Q2)-def}
\end{equation}
and $C  \equiv \ln (\alpha_{g_1}(Q_0^2)/\pi)$. 
To study how the complex structure of the DFs affects behavior of partons, we consider their propagators in Euclidean coordinate space. For the gluons,
a Wick rotation 
and Fourier transform provides the Euclidean propagator $D_E$ in coordinate space:
\begin{equation}
D_E(x) = \int \frac{d^4 Q}{(2\pi)^4}   e^{i Q  .  x}   D_E(Q^2), 
\label{eq:Euclidean_prop_main}
\end{equation}
where $x$ and $Q^2$ are now Euclidean 4-vectors, and $D_E(Q^2)$ is obtained by Wick-rotating Eq.~(\ref{eq:dressed-prop-gluons}).
The integration is  performed in the Supplementary material, Section~\ref{ap:gluon_prop_calc}. In the large $x$ limit (formally, $x\gg 1/Q$ in the Fourier transform), it yields:
\begin{equation}
D_E(x)= \frac{\mathcal{C}_0 ~ e^{-\frac{\Lambda_s}{\sqrt{2}}|x|}}{|x|^{\sfrac{5}{2}-d_a}}  
\cos \big(\theta(x) \big),
\label{eq:DE(x)_app_main}
\end{equation}
with the constants $\mathcal{C}_0$ and $d_a$ given by Eqs.~(\ref{eq:C0_g}) and (\ref{eq:gamma}), and $\theta(x)$  by Eq.~(\ref{eq:theta_g}). 
The $e^{-\frac{\Lambda_s}{\sqrt{2}}|x|}$ factor exponentially suppresses the gluon propagator 
in the IR with a characteristic length of $\sqrt{2}/\Lambda_s$ 
as expected if $\Lambda_s$  characterizes the confinement scale.
For a free field with a non-zero simple pole $Q_*$ (e.g. the field mass), $D_E(x) \propto  e^{-\sqrt{|Q_*^2|} |x|}/|x|^{3/2}$. 
In Eq.~(\ref{eq:Euclidean_prop_main}) however, the singularities of $D_E(Q^2)$ modify the free field case to $\sfrac{3}{2}+1-d_a$.
($d_a = \frac{4}{12+3\pi^2} \simeq 0.1$ is numerically small, contributing less than 4\% to the power.)
In all, the leading IR behavior of $D_E(x)$ depends on 
(I) the singularity positions $Q^2 = \pm i\Lambda_s^2$ and 
(II) the real part of the branch point power, $d_a$. 
The key factor, $e^{-\frac{\Lambda_s |x|}{\sqrt{2}}}$, however, comes solely from (I). The gluon propagator $D_E(x)$ is shown in Fig.~\ref{fig:prop} for 
$\Lambda_{g_1}=0.85$~GeV~\cite{Deur:2016tte,Deur:2023dzc} in the $g_1$ RS. (This RS value underlines that the asymptotic behavior (\ref{eq:DE(x)_app_main}) need not be 
the same as those found using different RS and gauges, especially since the present framework does not have ghost fields.)

The quark propagator calculation proceeds similarly to the gluon one and yields:
\begin{equation}
F_E(x)  =  \frac{\mathcal C'_0 e^{-\frac{\Lambda_s}{\sqrt{2}}|x|} }{|x|^{\sfrac{5}{2}-d_a}} 
\cos \big(\theta'(|x|) \big).
\label{eq:quark_prop_main}
\end{equation}
The derivation is provided in the Supplementary material, Section~\ref{ap:quark_prop_calc}, with $\mathcal C_0'$ and $\theta'(|x|)$ given by Eqs.~(\ref{eq:C0_q}) and (\ref{eq:theta_q}). 
$F_E(x)$ decays with the same scale $\sfrac{\Lambda_s}{\sqrt{2}}$ and $|x|^{-\sfrac{5}{2}-d_a}$ power as $D_E(x)$ .
That quarks and gluons have a single confinement scale comes from their DFs having and identical analytic structure:
$Z_2(Q^2) = c(Q^2)  Z_3(Q^2)$, Eq.~(\ref{eq:Z2}), with $c(Q^2)$ an analytic function. 
This in turn arises from assuming that at least one of the vertex DFs is analytical. 
Without this, the analytic structure could well be distributed differently among the DFs, implying different suppression factors, which would
be physically undesirable.


\section{Discussion and Conclusion}

We have shown how the analyticity properties of $\alpha_{g_1}$ imply that the $\ln (Z_i)$ are complex-valued functions with branch cuts in the complex $Q^2$ plane. 
This directly impacts parton propagation, damping it and thereby leading to color confinement without imposing it explicitly.
Assuming the analyticity of one of the vertex DFs makes the dissipative 
component of propagation to be
equally shared between quarks and gluons, neither sector playing a more important role.
Consequently, gluons and (light) quarks share the same propagator form,
$ P(x)= \frac{Ae^{-\frac{\Lambda_s}{\sqrt{2}}|x|}}{|x|^{\sfrac{5}{2}-d_a}} \Big[\cos \big(\frac{\Lambda_s|x|}{\sqrt{2}} - \frac{\pi}{2} d_a \ln|x| + B\big) \Big],$ 
where $A$ and $B$ are specific to the quark and gluon cases, e.g. $A \propto \slashed x/|x|$ for quarks, reflecting their spin-$\sfrac{1}{2}$ nature. 
To reach such results, we used a  local representation of $Z_3(Q^2)$ 
valid near the singularities $Q^2=\pm i\Lambda_s^2$  displayed by
$\alpha_{g_1}$ in the IR. Consequently, the behavior $e^{-\frac{\Lambda_s}{\sqrt{2}}|x|}  |x|^{-\sfrac{5}{2}+d_a}$ describes only the long-distance 
behavior of the propagators and not necessarily the pQCD regime, just like this latter yields a meaningless behavior for $\alpha_s$ below the Landau pole, namely a negative coupling that becomes infinite as it reaches that pole. 
The $e^{-\frac{\Lambda_s}{\sqrt{2}}|x|}$ factor exponentially suppresses the propagation of parton at large $x$ 
with a characteristic length of $\sqrt{2}/\Lambda_s$. Interestingly, $\Lambda_s/\sqrt{2}$ is remarkably close to the propagation-suppressing
emergent gluon mass $m_0=0.43(1)$~GeV 
obtained from lattice QCD and DSE in the MOM RS~\cite{Ding:2022ows}. Indeed, in this RS, $\Lambda_s=0.62$~GeV hence a scale $\Lambda_s/\sqrt{2}=0.44$~GeV. 

This approach provides an intuitive picture of confinement analogous to motion dissipation in classical physics. Whether we consider mechanical or electrical
systems, imaginary poles induce dissipative modes, damping the evolution of the system when poles have an imaginary component. 
For color confinement, the value of $\Lambda_s$ 
is such that the poles of $\alpha_s$ are indeed purely imaginary~\cite{deTeramond:2024ikl}, thereby maximally suppressing the spatial propagation of color particles. 
This picture also describes UV's asymptotic freedom: although $\Lambda_s^2$  
 is fixed and non-zero, having $Q^2 \gg \Lambda_s^2$ 
makes the scale to become more and more negligible, so propagation suppression by the imaginary singularities becomes less and less and as 
$Q^2 \to \infty$, $\kappa^2 \equiv  0$ and the partons evolve (propagate) freely. 

The general description of damping via complex singularities describes the evolution of the system without specifying the 
microscopic dynamical mechanism responsible for them. For example, a pole describing the damping of an RLC circuit does 
not specify the microscopic origin of electrical resistance, such as the scattering of conduction electrons from vibrating ions, 
structural defects, or impurities. Similarly, a pole describing friction in a mechanical system does not specify its microscopic 
origin in the interactions and deformation of the contacting surfaces. Likewise, the present description does not describe the 
underlying microscopic origin of the $\pm i\Lambda_s/\sqrt{2}$ singularities. Several microscopic mechanisms and different 
characterizations of confinement have been proposed~\cite{Greensite:2011zz}, including for the former the Gribov--Zwanziger 
framework~\cite{Gribov:1977wm, Zwanziger:1981kg, Vandersickel:2012tz} or dynamical mass generation through the 
Schwinger mechanism~\cite{Schwinger:1962tn,Schwinger:1962tp}, as studied in DSE and lattice QCD 
approaches~\cite{Ding:2022ows,Aguilar:2022thg} and the lack of asymptotic colored states or violation of reflection positivity 
of Schwinger functions for the latter. These can therefore be viewed as complementary: the present analysis describes an 
effective, global feature of color propagation, while functional and lattice approaches investigate its underlying dynamics.

~

\noindent {\bf Acknowledgments} We  thank S. J. Brodsky, G. F. de T\'eramond and C. D. Roberts for useful discussions. 
This material is based upon work supported by the U.S. Department of Energy, Office of Science, Office of Nuclear 
Physics under contract 89243126CSC000213. 

\bibliography{pole-flow-confinement}

\pagebreak

\appendix
\section{Supplementary material}

\subsection{Constraints on the phases $b_i$  and on vertex analyticities \label{Constraints on  b_i}}
Here, we show that if one of the vertex functions is analytical, say $Z_1$, then so is the other (say $Z_{3g}$).
We write $\ln Z_i(Q^2) \equiv a_i(Q^2) + i b_i(Q^2)$, where $a_i$ and $b_i$ are real functions and,
respectively, the log modulus and phase of the DFs. 
In Ref.~\cite{deTeramond:2024ikl}, it was shown that $\alpha_{g_1}$ obeys maximal analyticity, 
viz that the imaginary parts of its complex-conjugate poles are extremal, which occurs when the poles reach the imaginary axis (Fig.~\ref{fig:flow}).
This constraint can be implemented using Lagrange multipliers with a quadratic variational functional 
\begin{equation}
\mathcal{F}[Z_2,Z_3] =\int_0^\infty dQ^2  \big[2 b_2(Q^2)^2 + b_3(Q^2)^2 \big],
\label{eq:function-lagmult} 
\end{equation}
where the weights reflect the factors in Eq.~(\ref{eq:constraint}), and $C$
and $b_1$ are absent in (\ref{eq:function-lagmult}) because $C$ is real and $Z_1$ is analytic.
Indeed, a nonzero $b_1$ would produce a factor $(Q^2-Q_*^2)^{b_1}$ corresponding to a branch-point singularity. Thus, analyticity imposes $b_1=0$, with the resulting constant absorbed in $C$.
$\mathcal{F}$ is the simplest convex functional of $b_2$ and $b_3$ consistent with Eq.~(\ref{eq:constraint}), 
ensuring one has a single extremum. (In any case, one may show that any convex functional yields the same final result,  Eq.~(\ref{eq:b2b3})). 
Choice (\ref{eq:function-lagmult}) is confirmed by the fact that one recovers the result, Eq.~(\ref{eq:b2b3}), by solving Eqs.~(\ref{eq:a_s-qqg-vert}) and (\ref{eq:a_s-3g-vert}) once the vertices DFs are assumed to be analytic: $2b_2 + b_3 = 3b_3$, so $b_2=b_3$.
Maximal analyticity is reached for the DFs that extremize $\mathcal{F}$ under the constraint Eq.~(\ref{eq:constraint}). 
It can be splits into a constraint on the $a_i$, $2a_1  + 2a_2 + a_3 + C = \mathrm{Re}\big(I(Q^2)\big)$, and a constraint on the $b_i$:
\begin{align}
2b_2(Q^2) + b_3(Q^2) &= \mathrm{Im}\big(I(Q^2)\big). 
\label{eq:imag_constraint_qqg}
\end{align}
We implement Eq.~(\ref{eq:imag_constraint_qqg}) with the Lagrange multiplier $\lambda$,
$\tilde{\mathcal{F}} = \int_0^\infty dQ^2 \left[ 2 b_2^2 + b_3^2 - \lambda\left(\mathrm{Im}\big(I(Q^2)\big) - 2b_2 - b_3 \right) \right]$.
Then $\sfrac{\delta \tilde{\mathcal{F}}}{\delta b_2} =0$ 
and $\sfrac{\delta \tilde{\mathcal{F}}}{\delta b_3} = 0$ 
yield $b_2 = b_3$. Injecting this in
Eq.~(\ref{eq:imag_constraint_qqg}) then gives,  
\begin{equation}
3b_2 (Q^2)= \mathrm{Im}\big(I(Q^2)\big). 
\label{eq:imag_constraint_3g}
\end{equation}
In all,
\begin{equation}
b_2(Q^2) = b_3(Q^2) = \frac{1}{3} \mathrm{Im}\big(I(Q^2)\big).
\label{eq:b2b3}
\end{equation}
Thus, the phases of the quark and gluon DFs are the same as they evolve with $Q^2$.

The only assumption used here is that $Z_1$ is analytic. It implies that $Z_{3g}$ is analytic too since from Eq.~(\ref{eq:a_s-3g-vert}),  $3 b_3 + 2b_{3g}  = \mathrm{Im} ~I$. Then, with Eq.~(\ref{eq:b2b3}), 
$b_{3g} = 0$. 
Without assuming vertex analyticity, Eqs.~(\ref{eq:imag_constraint_qqg}) and (\ref{eq:imag_constraint_3g})  would be
\begin{align}
2 b_1(Q^2) + 2 b_2(Q^2) + b_3(Q^2)  &=  \mathrm{Im} \big(I(Q^2)\big) \label{eq:im1} \\
3 b_3(Q^2) + 2b_{3g}(Q^2) &=  \mathrm{Im} \big(I(Q^2)\big) \label{eq:im2}
\end{align}
that is
\begin{equation}
b_1(Q^2) +b_2(Q^2) = b_3(Q^2) + b_{3g}(Q^2),
\label{eq:identity}
\end{equation}
which must hold for all $Q^2$. The $b_1$, $b_2$, $b_3$, and $b_{3g}$ pertain to DFs from different sectors of QCD,
namely the quark-gluon vertex ($b_1$), the three-gluon vertex ($b_{3g}$), and the propagators ($b_2$ and $b_3$). 
No known Ward/Slavnov-Taylor identity, or symmetry enforces an exact relation between vertex and propagator phases of this cross-sector form. In fact, 
each DF obeys its own distinct equation within the coupled DSE tower. Satisfying Eq.~(\ref{eq:identity}) for every $Q^2$
without such a symmetry would thus require fine-tuning between distinct dynamical sectors. 
The simplest and most natural solution is $b_1 = b_{3g} = 0$, i.e., analytic vertex DFs. 
While this suggests the analyticity of vertex DFs, one cannot exclude that perhaps a dynamical mechanism or self-consistency condition
imposes Eq.~(\ref{eq:identity}) for $b_1\neq 0$ and $b_{3g} \neq 0$. It would be interesting to establish whether such a mechanism can exist.

\subsection{Gluon propagator \label{ap:gluon_prop_calc}}

Here, we provided the detailed integration of the Euclidean gluon propagator. It proceeds, in the large $|x|$ limit, in a standard manner
using the method of steepest descent~\cite{Arfken-Weber}.
For convenience, we repeat Eq.~(\ref{eq:Euclidean_prop_main}),
\begin{equation}
D_E(x) = \int \frac{d^4 Q}{(2\pi)^4}   e^{i Q  .  x}   D_E(Q^2)~~~~~~[\rm Eq.~(\ref{eq:Euclidean_prop_main})],
\label{eq:Euclidean_prop}
\end{equation}
and remind that $x$ and $Q^2$ are now Euclidean 4-vectors and $D_E(Q^2)= Z_3(-Q^2)/Q^2$ comes by Wick-rotating Eq.~(\ref{eq:dressed-prop-gluons}).
Since $D_E(Q^2)$ depends only on $Q^2$, then  $D_E(x)$ depends only on $|x|$ and is a 4D Fourier transform of 
the radially symmetric  $D_E(Q^2)$, viz $|x| D_E(|x|)= (2 \pi)^2 H\big(D_E(Q^2)\big)$, with $H$, the Hankel transform. This gives
\begin{equation}
D_E(|x|)=\frac{1}{(2\pi)^2 |x|} \int_0^\infty dQ  Q^2 J_1(Q|x|) D_E(Q^2).
\label{eq:spatial-gluon-prop}
\end{equation}
with $J_1$ the Bessel function of the first kind. The IR behavior of $D_E(|x|)$ is determined by the analytic structure of $D_E(Q^2)$. 
Rewriting $Q^2 \equiv s$ and changing the integration measure to $ds$, we obtain
\begin{equation}
D_E(|x|) =\frac{1}{(2\pi)^2 |x|} \int_0^\infty ds   \frac{\sqrt{s}}{2} J_1(\sqrt{s} |x|) D_E(s).
\label{eq:spatial-gluon-prop2}
\end{equation}
Using the Poincar\'e asymptotic (large $|x|$) expansion of the Bessel function  
\begin{equation}
J_1(\sqrt{s} |x|) = \frac{1}{\sqrt{2\pi \sqrt{s} |x|}} \left[e^{i\sqrt{s}|x| -i3\pi/4} + e^{-i\sqrt{s}|x| + i3\pi/4} \right]\Big[1+\mathcal O(\sfrac{1}{\sqrt{s}|x|})\Big], 
\label{eq:J1-representation}
\end{equation}
Eq.~(\ref{eq:spatial-gluon-prop2}) becomes at first order,
\begin{equation}
D_E(|x|) = \frac{1}{2(2\pi)^{\sfrac{5}{2}}|x|^{\sfrac{3}{2}}} \int_0^\infty ds ~s^{\sfrac{1}{4}} \Big( e^{i\sqrt{s}|x| -i3\pi/4} + e^{-i\sqrt{s}|x| +i3\pi/4} \Big) D_E(s).
\label{eq:spatial-gluon-prop3}
\end{equation}
The factor $|x|^{-\sfrac{3}{2}}$ is standard 
 for a massive (actual mass or effective mass as here) free field in the $|x| \gg Q^{-1}$ limit, 
 but we will see that the branch points of $D_E(s)$ modify the canonical \sfrac{3}{2} power.

In the asymptotic limit, the analytic structure of $D_E(s) = Z_3(-s)/s$ comes from the poles $s \equiv Q^2 = \pm i\Lambda_s^2$ of 
$\alpha_{g_1}$ in the IR. From Eq.~(\ref{eq:a_s-3g-vert}) and the analyticity of $Z_{3g}$, 
the poles of $\alpha_{g_1}$ correspond to the singularities of $Z_3$, so
\begin{equation}
sD_E(s) = Z_3(-s) = (-s - i\Lambda_s^2)^{\gamma} (-s + i\Lambda_s^2)^{\gamma}   \tilde h(s) = (s + i\Lambda_s^2)^{\gamma} (s - i\Lambda_s^2)^{\gamma} h(s),  
\label{eq:Z_3}
\end{equation}
where $\tilde h(x)$ is analytic and non-zero near these points, and $h(x) \equiv (-1)^{2\gamma} \tilde h(x)$. 
The value of $\gamma$ is determined by  the behavior of $I(Q^2)$, Eq.~(\ref{eq:I(Q2)-def}),
near $s = i\Lambda_s^2$ (or equivalently $s = - i\Lambda_s^2$). Expanding the denominator of the integrant of $I(s)$ around the pole, and using 
residues, we find  $I(s)=\int_0^{s} \frac{du}{f(u)} \simeq R\ln(s-i\Lambda_s^2) +a(s)$ near the pole, 
where $R= -1/(1 + i\frac{\pi}{2})$ is the residue of the integrant of $I(s)$, and $a(s)$ is an analytic function that need not be specified. 
Then, from Eq.~(\ref{eq:a_s-3g-vert}), $Z_3 = (s-i\Lambda_s^2)^{R/3} b(s)$, with $b(s)$ an analytic function.
Thus, the power in Eq.~(\ref{eq:Z_3}) is
\begin{equation}
\gamma = \frac{R}{3} = \frac{-1}{3 + i3\pi/2}= d_a(i\frac{\pi}{2}-1),
\label{eq:gamma}
\end{equation}
with $d_a \equiv \frac{4}{12+3\pi^2}$. [{\footnotesize Incidentally, if $\gamma=-1$, then $h(s) \propto s^2$ from dimensional considerations and (\ref{eq:Z_3}) becomes $D_E(s) \propto \frac{s}{[s+ i\Lambda_s^2][s- i\Lambda_s^2]}= \frac{s}{s^2 + \Lambda_s^4}$, which is Gribov's gluon propagator (e.g.,~\cite{Deur:2016tte}) with the Gribov IR scale identified to $\Lambda_s$. This propagator has simple poles rather than the more generic branch cuts. In fact, we do not expect the Gribov propagator to be equal to that discussed here since in Gribov's approach, the dynamics is controlled by Faddeev–Popov ghosts, which suppress the gluon propagator in the IR. Thus, in an approach without ghost like here, the gluon propagator may be different. Yet, that (\ref{eq:Z_3}) accommodates Gribov's propagator shows the present formalism generality.}]

Using $t \equiv s - {i\Lambda_s^2}$, Eq.~(\ref{eq:Z_3}) can be expanded so that 
$s^{\sfrac{1}{4}} D_E^{(+)}(s)=\frac{Z_3(-s)}{s^{\sfrac{3}{4}}} = \frac{t^{\gamma}}{(i\Lambda_s^2)^{\sfrac{3}{4}}} (2i\Lambda_s^2)^{\gamma} h({i\Lambda_s^2})\big(1-\mathcal O(i\Lambda_s^2 t)\big)$, where the $^{(+)}$ in $D_E^{(+)}(s)$ signals that we focus for now on the $+i\Lambda_s^2$ branch point. 
The factor $(2i\Lambda_s^2)^{\gamma}$ comes from the other singularity term since for $D_E^{(+)}(s)$, the integral strength comes from near $i\Lambda_s^2$ and the second singularity term contributes $\simeq (2i\Lambda_s^2)^{\gamma}$.
Likewise, we  expand the phase relevant to that branch point, 
$e^{i\sqrt{s}|x|} \simeq e^{i\sqrt{{i\Lambda_s^2}}|x|} e^{i \frac{t}{2\sqrt{{i\Lambda_s^2}}} |x|} \equiv e^{i\sqrt{{i\Lambda_s^2}}|x|} e^{i u} $, where 
$u \equiv \frac{t}{2\sqrt{{i\Lambda_s^2}}} |x|$. Using these expressions in Eq.~(\ref{eq:spatial-gluon-prop3}) yields:
\begin{equation}
D_E^{(+)}(|x|) =  \frac{(2i\Lambda_s^2)^{\gamma} e^{i\sqrt{{i\Lambda_s^2}}|x|}}{2(2\pi)^{\sfrac{5}{2}}|x|^{\sfrac{3}{2}}}   \frac{e^{-i3\pi/4}}{{(i\Lambda_s^2)}^{\sfrac{3}{4}}} 
\left(\frac{2\sqrt{{i\Lambda_s^2}}}{|x|}\right)^{1+\gamma} h({i\Lambda_s^2}) \int_0^\infty du   u^{\gamma} e^{i u}. \nonumber 
\end{equation}
Since $\int_0^\infty dv   v^{\gamma} e^{-v}=\Gamma \left(1+\gamma \right)$, 
then $\int_0^\infty du   u^{\gamma} e^{i u}=e^{i \frac{\pi}{2}(\gamma+1)}\Gamma \left(1+\gamma \right)$) and the above integrates to:
\begin{equation}
D_E^{(+)}(|x|) =  \frac{(2i\Lambda_s^2)^{\gamma} e^{i\sqrt{{i\Lambda_s^2}}|x|}}{2(2\pi)^{\sfrac{5}{2}}|x|^{\sfrac{3}{2}}}   \frac{e^{i\frac{\pi}{2}(\gamma-\sfrac{1}{2})}}{{(i\Lambda_s^2)}^{\sfrac{3}{4}}} 
\left(\frac{2\sqrt{{i\Lambda_s^2}}}{|x|}\right)^{1+\gamma} \Gamma \left(1+\gamma\right)h({i\Lambda_s^2}).
\label{eq:spatial-gluon-prop-plus0}
\end{equation}
Simplifying the various factors 
and since $\sqrt{s}=\sqrt{{-i\Lambda_s^2}} + \frac{t}{2\sqrt{{-i\Lambda_s^2}}} + \mathcal{O}(t)$,
\begin{equation}
D_E^{(+)}(|x|) =  \frac{(4\Lambda_s^3)^{\gamma} e^{-\frac{\Lambda_s|x|}{\sqrt{2} }}}{(2\pi)^{\sfrac{5}{2}} \Lambda_s^{\sfrac{1}{2}} |x|^{\sfrac{5}{2}+\gamma}}   
e^{i\big(\frac{\Lambda_s|x|}{\sqrt{2}}  + \gamma \frac{5\pi}{4} -\frac{3\pi}{8} \big)}
 \Gamma \left(1+\gamma\right)h({i\Lambda_s^2}).
\label{eq:spatial-gluon-prop-plus}
\end{equation}
The additional $|x|^{-(1+\gamma)}$ power comes from the non-analytic behavior near $i\Lambda_s^2$ and produces an overall $|x|^{-(\sfrac{5}{2}+\gamma)}$ dependance.

The $-i\Lambda_s^2$ branch point proceeds similarly.  Now, $t \equiv s + {i\Lambda_s^2}$ and expanding Eq.~(\ref{eq:Z_3}) gives 
$s^{\sfrac{1}{4}} D_E^{(-)}(s)=\frac{t^{\gamma}}{(-i\Lambda_s^2)^{\sfrac{3}{4}}} (-2i\Lambda_s^2)^{\gamma} h({-i\Lambda_s^2})\big(1+\mathcal O(i\Lambda_s^2 t)\big)$.
The phase relevant to $-i\Lambda_s^2$ is 
$e^{-i\sqrt{s}|x|} \simeq e^{-i\sqrt{{-i\Lambda_s^2}}|x|} e^{-i \frac{t}{2\sqrt{{-i\Lambda_s^2}}} |x|} \equiv e^{-i\sqrt{{-i\Lambda_s^2}}|x|} e^{-i u} $
where now, $u \equiv \frac{t}{2\sqrt{{-i\Lambda_s^2}}} |x|$. This gives
\begin{equation}
D_E^{(-)}(|x|) = \frac{(-2i\Lambda_s^2)^{\gamma} e^{-i\sqrt{{-i\Lambda_s^2}}|x|}}{2(2\pi)^{\sfrac{5}{2}}|x|^{\sfrac{3}{2}}}   \frac{e^{+i3\pi/4}}{{(-i\Lambda_s^2)}^{\sfrac{3}{4}}} 
\left(\frac{2\sqrt{{-i\Lambda_s^2}}}{|x|}\right)^{1+\gamma} h({-i\Lambda_s^2}) \int_0^\infty du   u^{\gamma} e^{-i u}, \nonumber 
\end{equation}
and in turn, 
\begin{equation}
D_E^{(-)}(|x|) = \frac{(4\Lambda_s^3)^{\gamma} e^{-\frac{\Lambda_s|x|}{\sqrt{2} }}}{(2\pi)^{\sfrac{5}{2}} \Lambda_s^{\sfrac{1}{2}} |x|^{\sfrac{5}{2}+\gamma}}   
e^{-i\big(\frac{\Lambda_s|x|}{\sqrt{2}}  + \gamma \frac{5\pi}{4} -\frac{3\pi}{8} \big)}
 \Gamma \left(1+\gamma\right)h({-i\Lambda_s^2}).
\label{eq:spatial-gluon-prop-minus}
\end{equation}
viz $D_E^{(-)}(|x|) = \Big(D_E^{(+)}(|x|)\Big)^*$, which agrees with 
the known fact a Euclidean propagator is a real function: $D_E(|x|)=D_E^{(+)}(|x|) + D_E^{(-)}(|x|) = 2$Re$\big(D_E^{(+)}(|x|)\big)$. Thus, 
\begin{equation}
D_E(|x|) = \frac{2 e^{-\frac{\Lambda_s|x|}{\sqrt{2} }}}{(2\pi)^{\sfrac{5}{2}} \Lambda_s^{\sfrac{1}{2}} |x|^{\sfrac{5}{2}}}   
\mathrm{Re} \left[ \bigg(\frac{4\Lambda_s^3}{|x|}\bigg)^{\gamma} e^{i\big(\frac{\Lambda_s|x|}{\sqrt{2}}  + \gamma \frac{5\pi}{4} -\frac{3\pi}{8} \big)} \Gamma \left(1+\gamma \right) h(i\Lambda_s^2)   \right].
\label{eq:spatial-gluon-prop4}
\end{equation}
Using Eq.~(\ref{eq:gamma}), we finally get
\begin{equation}
D_E(|x|)= \frac{\mathcal{C}_0 ~ e^{-\frac{\Lambda_s}{\sqrt{2}}|x|}}{|x|^{\sfrac{5}{2}-d_a}}  
\cos \big(\theta(|x|) \big),
\label{eq:DE(x)_app}
\end{equation}
with 
\begin{equation}
\mathcal{C}_0 = \frac{2}{(2\pi)^{\sfrac{5}{2}} \Lambda_s^{\sfrac{1}{2}} \big(4\Lambda_s^3\big)^{d_a}}   e^{-\frac{5\pi^2}{8}d_a}  \big| \Gamma \big(1+ d_a\big[i\frac{\pi}{2} -1 \big] \big) h(i\Lambda_s^2)\big|
\label{eq:C0_g}
\end{equation}
and 
\begin{equation}
\theta(|x|)=\frac{\Lambda_s}{\sqrt{2}}|x| + \frac{\pi}{2}\Big[\ln\Big( \frac{4\Lambda_s^3}{|x|} \Big) - \frac{5}{2} \Big]d_a  - \frac{3\pi}{8} +
{\rm arg}\left(\Gamma \big(1+ d_a\big[i\frac{\pi}{2}-1\big] \big) \right)  + {\rm arg}\big( h(i\Lambda_s^2)\big).
\label{eq:theta_g}
\end{equation}


\subsection{Quark propagator \label{ap:quark_prop_calc}}

Deriving the quark propagator $F_E(x)$ proceeds similarly to that of the gluon. Wick-rotating $F(Q^2)$, Eq.~(\ref{eq:dressed-prop-quarks})  yields 
\begin{equation}
F_E(Q^2) = Z_2(Q^2)  \frac{-i\slashed{Q} + m}{Q^2 + m^2}, 
\label{eq:quark_prop1}
\end{equation}
with $Q$ now a Euclidean 4-momentum, 
and there are no poles associated with the bare propagator $\propto 1/(Q^2 + m^2)$. In coordinate space, 
\begin{equation}
F_E(|x|) = \int \frac{d^4 Q}{(2\pi)^4}  F_E(Q^2) e^{iQ .  |x|}. \nonumber
\end{equation}
Since $Q_\mu e^{iQ.|x|} = -i \partial_\mu e^{iQ.  x}$, then $\int \frac{d^4Q}{(2\pi)^4}  \slashed{Q}  e^{iQ.|x|}   \frac{Z_2(Q^2) }{Q^2 + m^2}=- i \slashed{\partial}
\int \frac{d^4Q}{(2\pi)^4} e^{iQ.|x|}  \frac{Z_2(Q^2) }{Q^2 + m^2}$. Thus, 
$F_E(|x|) =\left(-i\slashed{\partial} + m\right)\big(  \int \frac{d^4 Q}{(2\pi)^4} e^{iQ .  |x|} \frac{Z_2(Q^2) }{Q^2 + m^2}  \big)$.
Since $Z_2(Q^2) /(Q^2 + m^2)$ depends only on $Q^2$, by rotational invariance, 
\begin{equation}
F_E(|x|) = \frac{-i\slashed{\partial} + m}{(2\pi)^2 |x|} \int_0^\infty dQ~Q^2 J_1(p|x|) \frac{Z_2(Q^2) }{Q^2 + m^2}.
\label{eq:quark_prop2}
\end{equation}
Hence, the spinor structure results from $(-i\slashed{\partial}+m)$ acting on a scalar function of $x$. 
Setting again $s \equiv Q^2$ and with $J_1(\sqrt{s}|x|)$ given by Eq.~(\ref{eq:J1-representation}), Eq.~(\ref{eq:quark_prop2}) becomes
\begin{equation}
F_E(|x|) \simeq \frac{-i\slashed{\partial} + m}{(2\pi)^{\sfrac{5}{2}} |x|^{\sfrac{3}{2}}} \int_0^\infty ds  \frac{s^{\sfrac{1}{4}}}{2} \left( e^{i\sqrt{s}|x|-i3\pi/4} + e^{-i\sqrt{s}|x|+i3\pi/4} \right) \frac{Z_2(Q^2) }{Q^2 + m^2}. 
\label{eq:quark_prop3}
\end{equation}
Equating Eqs.~(\ref{eq:a_s-qqg-vert}) and (\ref{eq:a_s-3g-vert}),
$Z_2(Q^2) = c(Q^2)  Z_3(Q^2)$ where $c(Q^2)$ is an analytic function. So, $Z_2$ and $Z_3$ have the same branch points $Q^2 = \pm i\Lambda_s^2$ 
which therefore determine the IR behaviors of both quark and gluon propagators.
Considering first the $Q^2 = i\Lambda_s^2$ branch point, we have
\begin{equation}
Z_2(Q^2) = c(Q^2)  Z_3(Q^2) = (Q^2 - i\Lambda_s^2)^{\gamma}(Q^2 + i\Lambda_s^2)^{\gamma} g(Q^2) ,  
\label{eq:Z2}
\end{equation}
with $g(Q^2) \equiv c(Q^2)h(Q^2)$,  analytic and non-vanishing near the branch points.
Focusing first on the contribution from near $s = i\Lambda_s^2$, we define $t \equiv s - i\Lambda_s^2$ and expand 
Eqs.~(\ref{eq:quark_prop1}) and (\ref{eq:Z2}) for $|t| \ll \Lambda_s^2$.
Inserting back $-i\slashed{\partial} + m$ in the integral (\ref{eq:quark_prop3}), we re-express it as 
$\slashed{Q} + m= \slashed{Q}^{(0)} + \gamma^\mu \delta Q_\mu +m$, with $(Q^{(0)})^2 = i\Lambda_s^2$. 
Near the singularity, $\gamma^\mu \delta Q_\mu \ll \slashed{Q}^{(0)}$ so to leading order, the momentum-space propagator is
\begin{equation}
F_E(t) \simeq \frac{t^{\gamma}(2i\Lambda_s^2)^{\gamma}}{i\Lambda_s^2 + m^2} \big[-i\slashed{Q}^{(0)} + m \big] g(i\Lambda_s^2).
\label{eq:quark_mom_prop}
\end{equation}

Near the singularity,  $\sqrt{s} = \sqrt{i\Lambda_s^2} + \frac{t}{2\sqrt{i\Lambda_s^2}} + \mathcal{O}(t^2)$ and thus
$e^{i\sqrt{s}|x|} \simeq e^{i\sqrt{i\Lambda_s^2}|x|} e^{ \frac{it}{2\sqrt{i\Lambda_s^2}} |x| }$. 
Using this,  $s^{\sfrac{1}{4}} \simeq (i\Lambda_s^2)^{\sfrac{1}{4}}$ and $u \equiv \frac{t}{2\sqrt{i\Lambda_s^2}} |x|$ as in the gluon case for its $i\Lambda_s^2$ singularity,  
the $i\Lambda_s^2$ contribution to Eq.~(\ref{eq:quark_prop3}) is
\begin{align}
F_E^{(+)}(|x|) \simeq & \frac{2^{\gamma}(i\Lambda_s^2)^{\gamma+1/4} e^{i(\sqrt{i\Lambda_s^2}|x|-\frac{3}{4}\pi)} } {2(2\pi)^{\sfrac{5}{2}} |x|^{\sfrac{3}{2}}} \frac{1}{i\Lambda_s^2 + m^2}g(i\Lambda_s^2)  \int_0^\infty dt ~ t^{\gamma} e^{\frac{it}{2\sqrt{i\Lambda_s^2}} |x|} \big[-i\slashed{Q}^{(0)} + m \big], \nonumber \\ 
= & \frac{2^{\gamma}(i\Lambda_s^2)^{\gamma+1/4} e^{i(\sqrt{i\Lambda_s^2}|x|-\frac{3}{4}\pi +  \frac{\pi}{2}(\gamma+1))} } {2(2\pi)^{\sfrac{5}{2}} |x|^{\sfrac{3}{2}}} \frac{\big[  i \sqrt{i\Lambda_s^2}\frac{\slashed{x}}{|x|} + m \big]}{i\Lambda_s^2 + m^2}  \left(\frac{2\sqrt{i\Lambda_s^2}}{|x|}\right)^{\gamma+1} g(i\Lambda_s^2) \Gamma  (1+\gamma).  \nonumber
\end{align}
where for the last line, we used $\int_0^\infty du   u^{\gamma} e^{i u}=e^{i \frac{\pi}{2}(\gamma+1)}\Gamma \left(1+\gamma \right)$,
and $\slashed{Q}^{(0)} = \sqrt{i\Lambda_s^2} \frac{\slashed{x}}{|x|}$ near the singularity, which follows from rotational invariance
under angular integration and that $x_\mu$ is the only available vector in coordinate space. 
Simplifying,
\begin{equation}
F_E^{(+)}(|x|) \simeq  \frac{2^{2\gamma+1}(i\Lambda_s^2)^{2\gamma+5/4} e^{i(\sqrt{i\Lambda_s^2}|x|-\frac{3}{4}\pi +  \frac{\pi}{2}(\gamma+1))} } 
{2(2\pi)^{\sfrac{5}{2}} |x|^{\sfrac{5}{2}+\gamma} }
\frac{\big[  i \sqrt{i\Lambda_s^2}\frac{\slashed{x}}{|x|} + m \big]}{i\Lambda_s^2 + m^2}g(i\Lambda_s^2) \Gamma  (1+\gamma). \nonumber
\end{equation}

The contribution $F_E^{(-)}$ from the other singularity, $-i\Lambda_s^2$, is the complex conjugate of $F_E^{(+)}(|x|)$,
and their sum yields the coordinate-space quark propagator $F_E(|x|) = F_E^{(+)}(|x|) + F_E^{(-)}(|x|)= 2 \mathrm{Re} F_E^{(+)}(|x|)$:
\begin{align}
F_E(|x|)  \simeq \frac{\mathcal C'_0 e^{-\frac{\Lambda_s}{\sqrt{2}}|x|} }{|x|^{\sfrac{5}{2}-d_a}} \cos \big(\theta'(|x|) \big)
\label{eq:quark_prop4}
\end{align}
which is the same form  as the gluon propagator but with different normalization
\begin{equation}
\mathcal C_0' = \frac{1} 
{2^{2d_a-1}\Lambda_s^{2d_a-5/2}  e^{\frac{\pi^2 d_a}{4}-\frac{\pi}{2}} (2\pi)^{\sfrac{5}{2}} [\Lambda_s^4+m^4]} \bigg[ ( m^2 - \Lambda_s^2)\frac{\Lambda_s}{\sqrt{2}} \frac{\slashed{x}}{|x|} + m^3 \bigg] 
\big| \Gamma  \big(1+d_a(i\frac{\pi}{2}-1)\big) ~ g(i\Lambda_s^2) \big| ,
\label{eq:C0_q}
\end{equation}
and different $x$-dependent phase
\begin{equation}
\theta'(|x|)=\frac{\Lambda_s}{\sqrt{2}}|x| +\frac{\pi}{2} \big(\ln|x| - 3\big)d_a  - \frac{\pi}{8} -\pi \ln (2) +\pi^2 + {\rm arg} \Big( \Gamma  \big(1+d_a(i\frac{\pi}{2}-1)\big) \Big) + {\rm arg} \big(g(i\Lambda_s^2) \big).
\label{eq:theta_q}
\end{equation}

\end{document}